\documentclass[twocolumn,twoside,slac]{revtex4}
\usepackage{graphicx}
\usepackage{fancyhdr}
\usepackage{hyperref}
\begin{document}

\title{The use of Ethernet in the DataFlow of the ATLAS Trigger \& DAQ}

%

\author{Stefan Stancu} 
\affiliation{CERN, Geneva, Switzerland and UPB Bucure\c{s}ti, Romania}
\author{Bob Dobinson} \affiliation{CERN, Geneva, Switzerland}
\author{Matei Ciobotaru} 
\affiliation{CERN, Geneva, Switzerland and UPB Bucure\c{s}ti, Romania}
\author{Krzysztof Korcyl} \affiliation{INP Cracow, Poland}
\author{Emil Knezo} 
\affiliation{CERN, Geneva, Switzerland and SAI, Eindhoven University of
Technology, Eindhoven, The Netherlands}
\author{On behalf of the ATLAS TDAQ DataFlow community\cite{DF}} 
\noaffiliation

\begin{abstract}
The article analyzes a proposed network topology for the ATLAS DAQ DataFlow,
and identifies the Ethernet features required for a proper operation of the
network: MAC address table size, switch performance in terms of throughput and
latency, the use of Flow Control, Virtual LANs and Quality of Service. We
investigate these features on some Ethernet switches, and
conclude on their usefulness for the ATLAS DataFlow network.
\end{abstract}

\maketitle

\thispagestyle{fancy}


\section{Introduction \label{sec:Intro}}
ATLAS is one of the five experiments foreseen to run
on the LHC (Large Hadron Collider) which is currently being built at CERN. The
proton-proton bunch crossings occur in ATLAS at approximately 40 MHz, the
detector recording around 2 MBytes of data per \textit{event}\footnote{The
data acquired from one bunch crossing represents an event.}. The amount of
raw data generated by the detector is extremely large: 80 TBytes/second.
However the final data rate which must be recorded to mass storage is of a
few tens of MBytes/second.

The ATLAS TDAQ (Triggered Data Acquisition) system selects the interesting
events  using a three layer trigger architecture: LVL1 (level 1), LVL2
(level2), and the EF (Event Filter). The LVL1 trigger is entirely build in
hardware, while LVL2 and the EF are implemented using PC farms. The output
rate of LVL1 can reach 75 KHz, while the LVL2 output rate is around 2 KHz.

This paper presents results of investigations into the Ethernet features
required for a proper operation of a network based TDAQ system, which reads
data from the Level 1 trigger output, runs LVL2 algorithms and supplies the
validated events to the Event Filter.

\section{ATLAS TDAQ Network-Based Architecture \label{sec:AtlsNetArch}}
\begin{figure*}
    \centering
    \includegraphics[width=135mm]{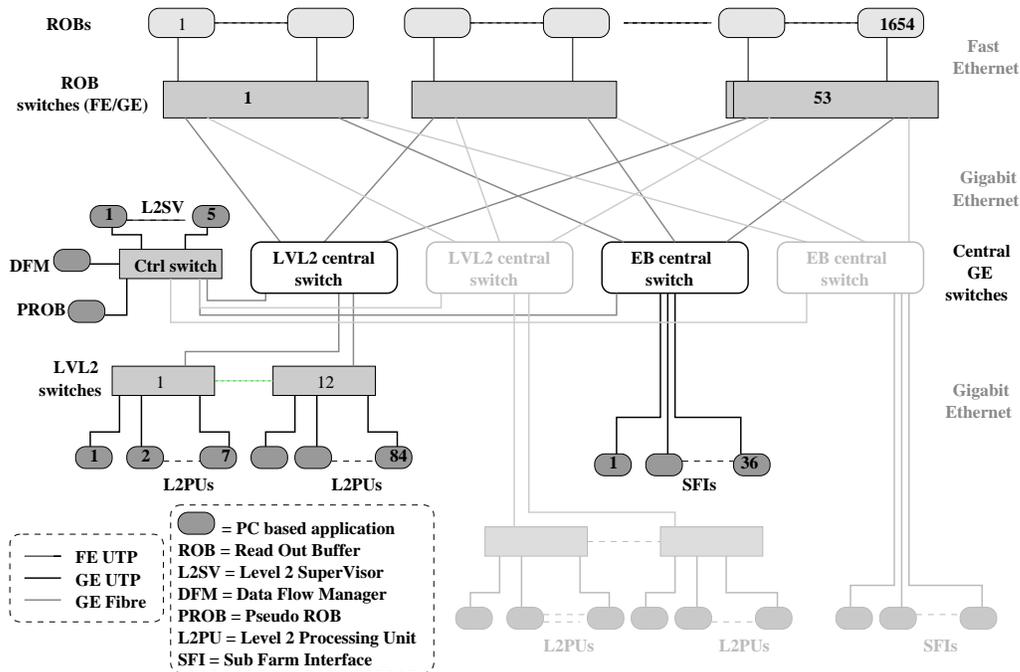}
    \caption{ATLAS Network-Based Architecture. }
    \label{fig:AtlsNetArch}
\end{figure*}

The ATLAS TDAQ Network-Based Architecture \cite{HP03} is presented in
Figure~\ref{fig:AtlsNetArch}. All the nodes except the ROBs (Read Out Buffers)
are PCs. The ROBs interface to the network is FE UTP (Fast Ethernet,
Unshielded Twisted Pair), while GE (Gigabit Ethernet) UTP  is used for PCs.
The network switches are interconnected through GE optical fibre links.

The events selected by the LVL1 trigger are buffered in approximately 1600 ROBs
(Read Out Buffers) at a rate up to 75 KHz. The DataFlow system (everything
below the ROBs in Figure~\ref{fig:AtlsNetArch}) receives RoI (Region of
Interest) information from the LVL1 trigger. The RoI points to the subset of
event data which led to the level 1 accept decision.  The DataFlow system plays a
double role:
\begin{itemize}
    \item it further investigates the events selected by the level 1 trigger
    and accepts/rejects them. This is the \textit{LVL2} (level 2) part.
    \item it gathers up the events which are validated by the LVL2 analysis.
    This is the \textit{EB} (Event Building) part.
\end{itemize}

\subsection{Message Flow \label{sec:AtlsNetArchMsgFlow}}
The message flow in the DataFlow system (see \cite{DC-012} for details) is
initiated by the L2SV (Level 2 Supervisor) which receives the RoI information
from the LVL1 trigger. The L2SV has the role of load balancing the event
processing task among the L2PUs (Level 2 Processing Units). The L2SV forwards
the RoI information to a L2PU having enough free resources. The L2PU algorithms
are incremental. It successively requests RoI information from the ROBs for
analysis until it reaches a decision. This process is called RoI Collection.
Once the event is validated/rejected the L2PU communicates its decision to the
L2SV. A detailed record of the validated event's analysis is passed to the PROB
(Pseudo ROB). The L2SV forwards the level 2 result to the DFM (Data Flow
Manager). This is the end of the LVL2 part, and the rest of the message flow
describes the Event Building.

If the LVL2 decision has been favorable, the DFM assigns an SFI (Sub Farm
Input) to gather up the event data. Upon reception of the event identifier,
the SFI requests the event data from all the ROBs (including the PROB), and
builds up a full event. Once all the fragments of the event have been
assembled, the SFI signals the end of event building to the DFM.  The
identifiers of the rejected events, as well as those of the completed ones,
are grouped and sent via a multicast message to all the ROBs (including the
PROB). Upon reception of the ``clear'' message the ROBs erase the specified
events from their memory. 

Multicast and broadcast handling and performance are switch vendor-specific
and users have no control over them. This is why the only multicast is the
clear message sent by the DFM to the ROBs. As the event identifiers are
grouped, the rate of this message is low (approx 300 Hz) and can be handled by
almost any Ethernet switch. All the other messages are unicasts. Moreover the
use of unicasts gives full control over the traffic patterns flowing through
the DataFlow network. 

The SFI buffers the completed events and subsequently passes them to the Event
Filter via a second interface. From the Event Filter point of view the SFI
acts like a server delivering events on demand\footnote{The architecture of
the Event Filter is outside the scope of this paper. A proposed architecture
can be found in \cite{HP03}.}.

\subsection{Network architecture \label{sec:AtlsNetArchNetArch}}

The DataFlow network is based on Ethernet technology. In the design
process \cite{HP03} care has been taken to keep the load on all Ethernet links
well below full capacity, in order to avoid congestion.

\begin{figure}
    \centering
    \includegraphics[width=\columnwidth]{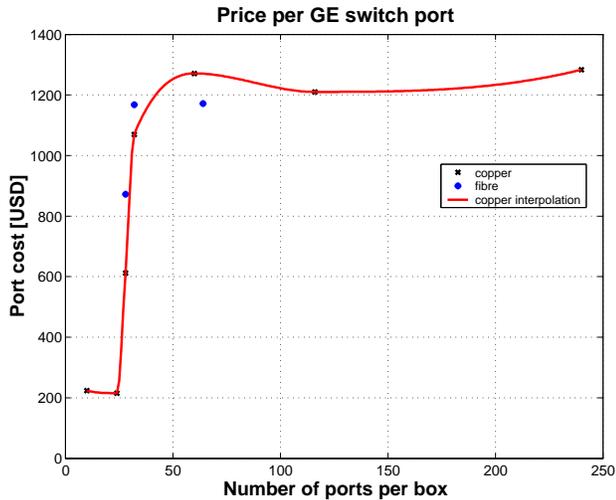}
    \caption{Gigabit Ethernet switch port price as function of the switch
    size.}
    \label{fig:SwPortPrice}
\end{figure}

For a better bandwidth utilization the components having a low throughput can
be grouped in clusters, using switches with a small number of FE or GE UTP ports
and fibre GE up-links.  Such concentrating switches are used for the ROBs and
the L2PUs. The grouping reduces the size of the central switches significantly.
The cost of the whole system goes down, as the ``per port'' price of small
switches is at the moment much lower than that of large switches (see
Figure~\ref{fig:SwPortPrice}).

The core of the network are the central switches (GE switches). The maximum
number of central switches is dictated by the number of up-links of the ROB
concentrating switches. The use of several central switches improves the fault
tolerance of the network, and allows the implementation of various staging
scenarios. The system shown in Figure~\ref{fig:AtlsNetArch} operates at full
capacity when both its stages are active (the second stage is drawn in
light colour). However, it can work at a lower rate using only one stage.

A stage contains two central switches: one for LVL2, and one for Event
Building. The separation of the two data flows at an early stage can improve event
processing latency for level 2 (replies to the LVL2 requesting nodes are not
mixed with fragments directed to the EB), and aids the implementation of
various Event Building traffic shaping schemes aimed at minimizing the packet
loss probability .

\subsection{Ethernet issues \label{sec:AtlsNetArchEthFeatures}}

For a proper operation of the ATLAS Baseline Architecture we need to
investigate a series of Ethernet features:
\begin{itemize}
    \item switches performance: throughput, packet loss, latency, MAC (Media
    Access Control) address table size;
    \item Flow Control behaviour at different levels of the network;
    \item VLAN (Virtual Local Area Network) implementation;
    \item QoS (Quality of Service);
    \item broadcast and multicast handling.
\end{itemize}
The results obtained from investigating the above mentioned features, as well
as their implication to the TDAQ system are the object of this paper, and will be
presented in section~\ref{sec:Measurements}.

\subsubsection{Why Ethernet?}
\begin{figure}
    \centering
    \includegraphics[width=\columnwidth]{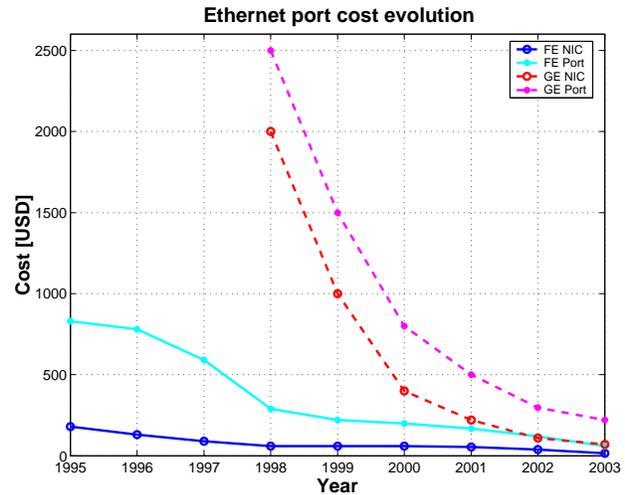}
    \caption{Port cost evolution for Fast Ethernet and Gigabit Ethernet }
    \label{fig:PortCostEvol}
\end{figure}

The reasons for using Ethernet technology for the DataFlow network are
its high performance and the low price (see Figure~\ref{fig:PortCostEvol}).
Ethernet products are now commodity.  The technology is multi-vendor, and we
foresee long term support for it.

The biggest step forward in the Ethernet evolution was the transition from
half duplex CSMA/CD (Carrier Sense Media Access with Collision Detection) to
full duplex. On a full duplex switched Ethernet network the bandwidth of each
line is full time guarantied, allowing a successful use of Ethernet in
real-time systems (as opposed to CSMA/CD, where the communication media is
shared and access to it is not guaranteed).

Moreover Ethernet has an evolutionary upgrade path to high speed. The 10 GE
(Gigabit Ethernet) IEEE standard was approved in 2002. For the moment the
price of 10 GE Ethernet switches is high. However the expected price drop over
the next few years would make them good candidates for the central switches.


\section{Ethernet features investigation \label{sec:Measurements}} 
We have used our customized traffic generators (see \cite{BAR02} and
\cite{DOB01}) for testing the Ethernet Features required by the ATLAS
Network-Based Architecture. The FE (Fast Ethernet) tester is a custom built
board, designed at CERN. It implements 32 FE ports (full-duplex 100 Mbps) using
Altera Flex FPGAs programmed in Handel-C. The GE tester is based on the Alteon
Gigabit Ethernet NIC (Network Interface Card). The card uses the Tigon II PCI
Ethernet Controller, which contains two customized MIPS CPUs, allowing a
flexible reprogramming. We have 4 FE boards (thus 128 FE ports), and around 30
Alteon NICs. 

The testers are capable of generating traffic with different packet sizes,
either with a specified CBR (Constant Bit Rate) or with a Poisson (negative
exponential) distributed inter-packet gap. They measure packet loss
and latency (300 ns precision), and also histogram the latencies (jitter) on a
\textit{per packet basis}. For more details about the testers consult \cite{BAR02}.

We have used this equipment to test both \textit{concentrating} and
\textit{central} switches from different manufacturers. By \textit{concentrating
switches} we denote switches with at least 2 GE optical fibre up-links and
either many FE UTP (Unshielded Twisted Pair) or several GE UTP ports. The
\textit{central switches} have a large number of GE optical fibre ports (at
least 30).

\subsection{Basic measurements on switches}
Switches are the key element of the ATLAS TDAQ Network-Based Architecture (see
section \ref{sec:AtlsNetArchNetArch}). They must
meet the throughput requirements of the architecture with a minimum latency
and packet loss.

Packet loss has a great penalty on the system's performance, as it implies time
outs and retries at the application level, which are time consuming. The
switches drop frames when their buffers overflow, therefore the bigger the
buffers, the smaller the probability of packet loss. The Ethernet Flow
Control helps preventing buffer overflow, but it does not solve the packet
loss problem completely (see section \ref{sec:MeasFC}).

We measure \textit{packet loss} and \textit{latency} for different Ethernet frame
sizes, different loads (from 10\% till 100\% of the line speed), with CBR
or with random (Poisson) inter-packet gap, using unicast, multicast and
broadcast traffic. 

The multicast  and broadcast tests proved the handling of such traffic is
vendor specific, and sometimes the maximum rate (throughput) is surprisingly
low (less than 10\% of the line speed). This is one of the reasons for choosing
a preponderant request-response message flow scenario (unicast traffic) for the
ATLAS Baseline Architecture (see section \ref{sec:AtlsNetArchMsgFlow}).

\begin{figure}
    \centering
    \includegraphics[width=\columnwidth]{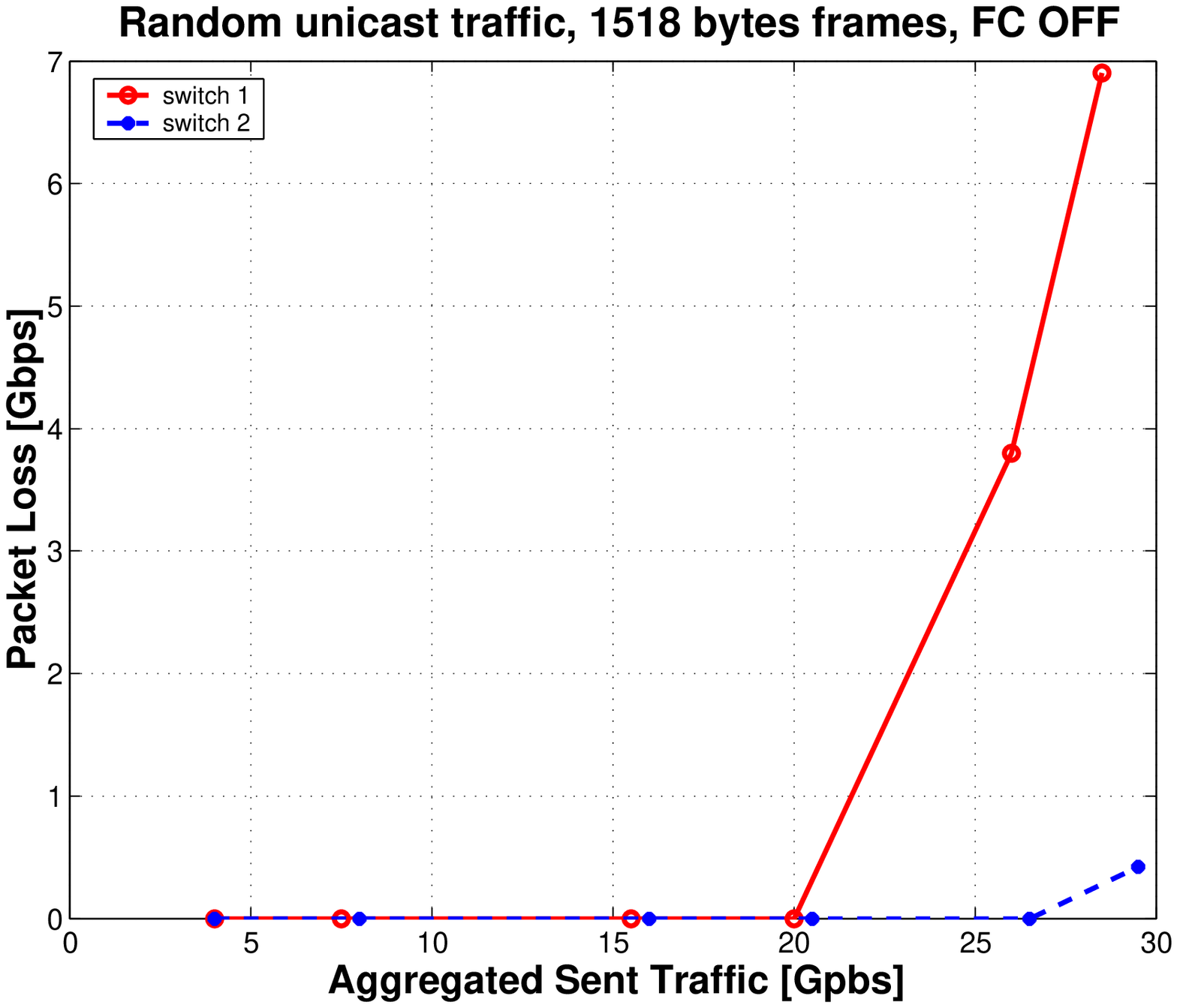}
    (a)
    \includegraphics[width=\columnwidth]{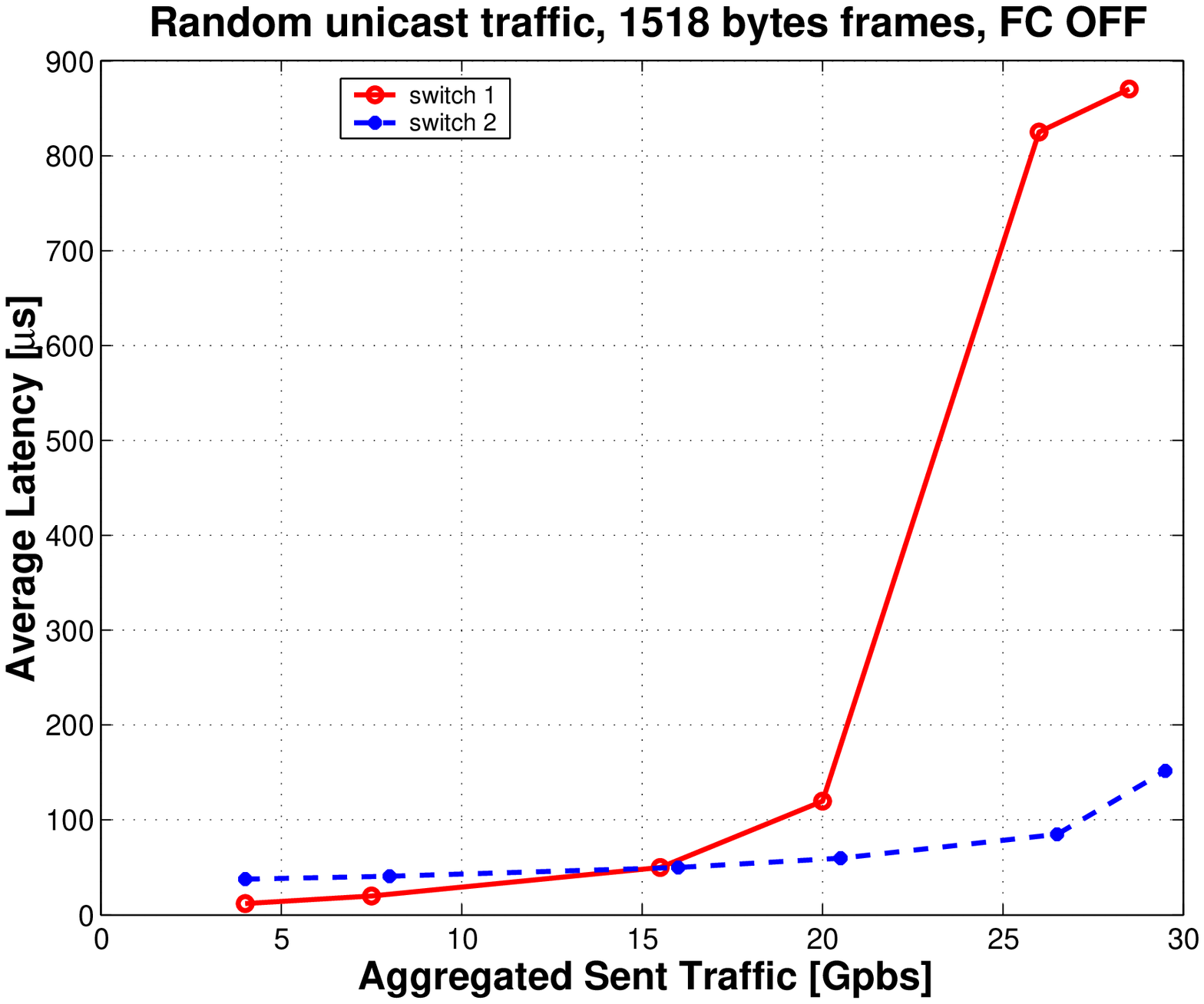}
    (b)
    \caption{Switch measurements for unicast traffic, Poisson inter-packet gap, 1518 bytes
    frames: (a) Packet loss (b) Average latency.}
    \label{fig:SwMeasurement}
\end{figure}

Figure~\ref{fig:SwMeasurement} shows the results from a test performed on two
different central switches, using 1518 bytes frames. We use 30 GE ports, each
one sending unicast traffic to all the others with a
negative exponential inter-packet gap. \textit{Switch 1}
becomes saturated when the offered load exceeds 66\% of the line speed. 
We first notice a slight increase of the latency followed by packet loss and a
significant growth of the latency once the buffers become full. \textit{Switch
2} handles the traffic much better (almost line speed). All the switches from
the ATLAS DataFlow network must operate \textit{below} the saturation
point, as packet loss and high latency values significantly diminish the
overall system performance. 

\subsection{MAC address table size}
The switch MAC address table contains the correspondence between the Ethernet
MAC addresses and the switch's ports associated to them. If the Ethernet
destination address from the header of a received frame is not known by the
switch, the frame will be forwarded through all its active ports
(\textit{flooding}). The DataFlow network has a large number of nodes (in
excess to 2000), which must be memorized by the switches in order to minimize
the flooding effect.

There are two types of MAC address table entries: \textit{static} and
\textit{dynamic}.  \textit{Static} entries are entered using a switch
management tool. This ensures that the switch will never flood frames.
However, if the network topology changes the user needs to update the MAC
address table. This is a great inconvenience for networks with large number of
nodes.  On the other hand no human intervention is needed if we use
\textit{dynamic} entries.  When a frame is received, the switch looks at its
Ethernet source address. If the address is unknown to the switch, it is added
to the MAC address table.  If the network topology changes and the same MAC
address is received on a different port, the MAC table will be updated
correspondingly. When the switch no longer receives frames from an address for
a certain amount of time (\textit{aging time}), it will erase that entry from
the MAC address table. A typical value for the aging time is some hundreds of
seconds. This is another motivation for the request response traffic. We want
to operate the DataFlow with dynamic MAC address table entries, and the
request-reply scenario minimizes the probability of aging.

In order to test the MAC address table size we have modified the Alteon traffic
generators. We have created a client which can generate up to 4096 MAC
addresses with different patterns. This client sends requests to the switch.
Two more NICs are set in promiscuous mode (they will receive all the frames
arriving to them, regardless of their Ethernet destination address).  One of
them is used as a server which responds using the destination address from the
received request as its own Ethernet source address.  This ``server'' will
emulate any number of Ethernet nodes. The second NIC which is set in
promiscuous mode is used as a ``listener'' for flooding.  We first perform a
\textit{learning} phase: after clearing the switch MAC address table, we
generate 4096 requests with different Ethernet destination addresses. The
switch does not know anything about them so it will flood them on all the
ports. Both the listener and the server will see them all. The server replies
will allow the switch to learn the injected Ethernet addresses.  Subsequently
we run the \textit{measurement phase}: we clear the counters both on the server
and the listener, and then repeat the generation of the same 4096 requests.
Let $N$ denote the number of frames received by the listener node during the
\textit{measurement phase}.  The listener should see only the flooded frames,
i.e the ones corresponding to the addresses not learned by the switch.
Therefore we are sure that the switch can accommodate $4096-N$ MAC addresses in
its table.

We have applied this method for different MAC address patterns on several
switches. Table \ref{tab:MACAddr} summarizes the results from a switch with
a peculiar behaviour. If we linearly increase the lower two bytes of the MAC
address, while keeping the others fixed (line 1) the switch learns 4096.
The same thing happens when we randomly generate the lower three bytes
(line 4), or when we mix the linear and random patterns (line 5).
The unexpected behaviour is revealed when we linearly increase either the
fourth and the fifth bytes (line 2), or the third and the fourth (line 3): the
switch MAC address table size is less than 80.

\begin{table}
\begin{center}
\caption{MAC address table measurements. \textit{xx}, \textit{yy} and
\textit{zz} are chosen arbitrarily but remain fixed. $\alpha\alpha$ are
linearly generated numbers, while $\beta\beta$ are random generated numbers.}
\label{tab:MACAddr}
\begin{tabular}{|l|r|}
    \hline
    \textbf{MAC address pattern}  & \textbf{MAC address table size}	\\
    \hline
    00:xx:yy:zz:$\alpha\alpha:\alpha\alpha$  & 4096 \\
    \hline
    00:xx:yy:$\alpha\alpha:\alpha\alpha$:zz  & 70 \\
    \hline
    00:xx:$\alpha\alpha:\alpha\alpha$:yy:zz  & 80 \\
    \hline
    00:xx:yy:$\beta\beta:\beta\beta:\beta\beta$  & 4096\\ 
    \hline
    00:xx:yy:$\beta\beta:\beta\beta:\beta\beta$	& 4096\\ 
    00:xx:yy:zz:$\alpha\alpha:\alpha\alpha$		& \\ 
    \hline
\end{tabular}
\end{center}
\end{table}

All the nodes from the DataFlow network except the ROBs are interfaced
via standard NICs. The MAC address of a NIC has the manufacturer code
reflected in the first four bytes, while the last two bytes of the address are
most likely random. This address pattern (line 4 from table \ref{tab:MACAddr})
causes no problem to the switch.
The ROBs are custom build hardware, and we have the freedom of choosing their
HW addresses. It is natural to linearly increase two of the bytes from the MAC
address. Our measurements show that for a proper operation of this particular
switch we are forced to vary the lower two bytes of the MAC address. When
choosing the ROB MAC addresses we must also make sure the first four bytes do not
coincide with any manufacturer code.

\subsubsection{MAC address aging}
The MAC address aging time should be large enough to avoid flooding once the
DataFlow system is operating. In the request-response message flow scenario
all the nodes send messages to the network periodically. At the level of ROB
switches and Central switches the SFI Event Building has the lowest rate
(around 30 Hz), which imposes an aging time larger than 30 ms. The aging time
in the switches we have tested is typically some hundreds of seconds,
therefore more than enough at this level of the network. Yet flooding may
potentially occur in the L2PU switches. These switches never forget the
addresses of the L2PUs (which request with an average rate of 10KHz), but they
may forget the addresses of some ROBs. L2PUs receive data only from a fraction
of the ROBs and there is a low probability that an L2PU cluster does not
request data from a given ROB for a period larger than the aging time. This
probability is fairly low, and the flooding has a small impact, as it will
most likely be confined to that concentrating switch.

Thus a typical value of hundreds of seconds for the MAC address table
\textit{aging time} assures a proper packet switching (no flooding after the
beginning of the run) in the DataFlow network.


\subsection{VLANs -- IEEE 802.1Q}
The extended header of the Ethernet frame may include a VLAN (Virtual LAN\footnote{Local
Area Network}) tag immediately after the Ethernet addresses. This
tag contains the VLAN ID (identifier), and also a priority field. The VLAN ID
allows more logical (virtual) LANs to coexist on the same physical LAN, while
the priority field allows the layer two traffic to be classified.
VLANs are crucial for the ATLAS Network-Based Architecture, as they assure a
loop free topology. In addition they bound the multicast/broadcast messages,
and provide QoS (Quality of Service) support.

\subsubsection{Loops and the Spanning Tree Protocol.}
The network topology of the ATLAS Network-Based Architecture
(Figure~\ref{fig:AtlsNetArch}) contains loops. An example of a loop is the
LVL2 central switch -- a ROB concentrating switch -- the EB central switch --
another ROB concentrating switch. Ethernet loops are illegal because they
perturb the MAC address tables for unicast traffic and they keep sending
forever multicasts and broadcasts (broadcast storms). The STP (Spanning Tree
Protocol) is designed to cut off the redundant links from a LAN in order to
maintain a loop free topology. We plan to use VLANs for maintaining a loop
free topology: a LVL2 VLAN (for the level 2 traffic) and an EB VLAN (for the
Event Building data flow). Several nodes need to be part of both VLANs: the
DFM, the PROB and the ROBs.

\begin{figure}
    \centering
    \includegraphics[width=70mm]{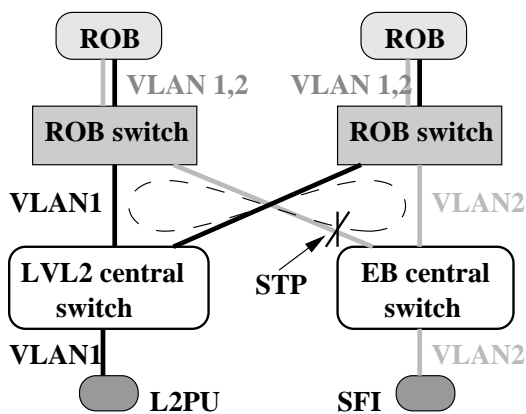}
    \caption{VLAN Ethernet loop setup. The potential loop appears in dashed
    line.}
    \label{fig:VLANLoops}
\end{figure}

The setup described in Figure~\ref{fig:VLANLoops} allows us to verify that
VLANs eliminate illegal loops, and also to check if the STP is aware of VLANs.
The tests performed showed that VLANs provide a loop free topology when the
STP is disabled. When the STP was active, it disabled one of the links in the loop.
This is due to the fact that the STP is not implemented per VLAN on
those switches.

We have full control over the DataFlow network. Therefore, if the STP is not
implemented per VLAN we can disable it, and carefully use VLANs for
maintaining a loop free topology.

\subsubsection{Traffic containment}
VLANs keep flooding, multicast and broadcast traffic inside their
defined bounds. To verify this feature we use one transmitter and several
receivers. The transmitter sends traffic to the investigated VLAN (unicast to
an address which is not known by the switch, multicast or broadcast). We place
one receiver in every VLAN which is defined on the switch, plus an additional
receiver outside any defined VLAN (i.e in the switch's default VLAN).  All
receivers outside the VLAN into which we inject traffic should receive no
frames.

We have carried out such tests for several switches in all of the following
situations: one VLAN, two VLANs with no shared ports and two VLANs with shared
ports. In no case were the VLAN boundaries crossed.

This feature is useful as we can restrict the number of nodes receiving a
multicast/broadcast message. For example, if we define a third VLAN in the
DataFlow network, containing the DFM and the ROBs, the clear message
multicasted by the DFM will be forwarded only to the nodes it should
reach (i.e. the ROBs).

\subsubsection{Partitioning}
VLANs divide a physical LAN into more logical ones. By analogy we want to
see if VLANs can partition one physical switch in two logical ones,
and also quantify the interference between the two partitions.

\begin{figure}
    \centering
    \includegraphics[width=70mm]{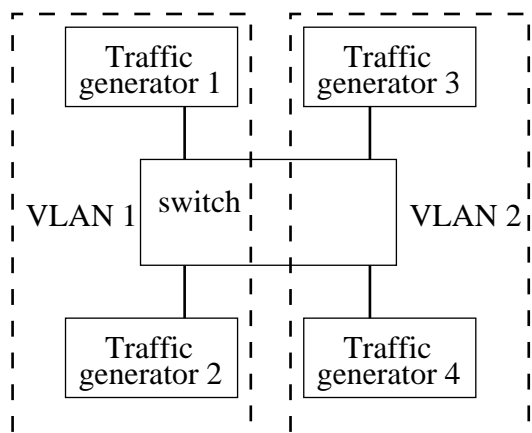}
    \caption{VLANs: switch partitioning setup.}
    \label{fig:VLANPart}
\end{figure}

In the setup presented in Figure~\ref{fig:VLANPart} we send independent
traffic into each of the two VLANs which do not share ports. For several
traffic loads in VLAN2 we measure the behaviour inside VLAN1. For the switch
we have tested the performance inside VLAN1 is insensitive to the amount of
traffic flowing through VLAN2. Therefore this switch can be partitioned in a
satisfactory way using VLANs.  This feature is highly dependent on the switch
architecture, and no assumption should be made a priori for any switch.

\subsubsection{Quality of Service (QoS)}
Quality of Service support is provided by the priority field from the VLAN tag.
Up to eight different priorities can be assigned to the frames, as the priority
field is three bits wide.  Switches may adopt different QoS schemes such as
\textit{strict priority} or \textit{weighted round robin} (WRR). If the line is
oversubscribed the latter algorithm allocates bandwidth for all the priorities
proportionally to their associated weight. 

\begin{figure}
    \centering
    \includegraphics[width=\columnwidth]{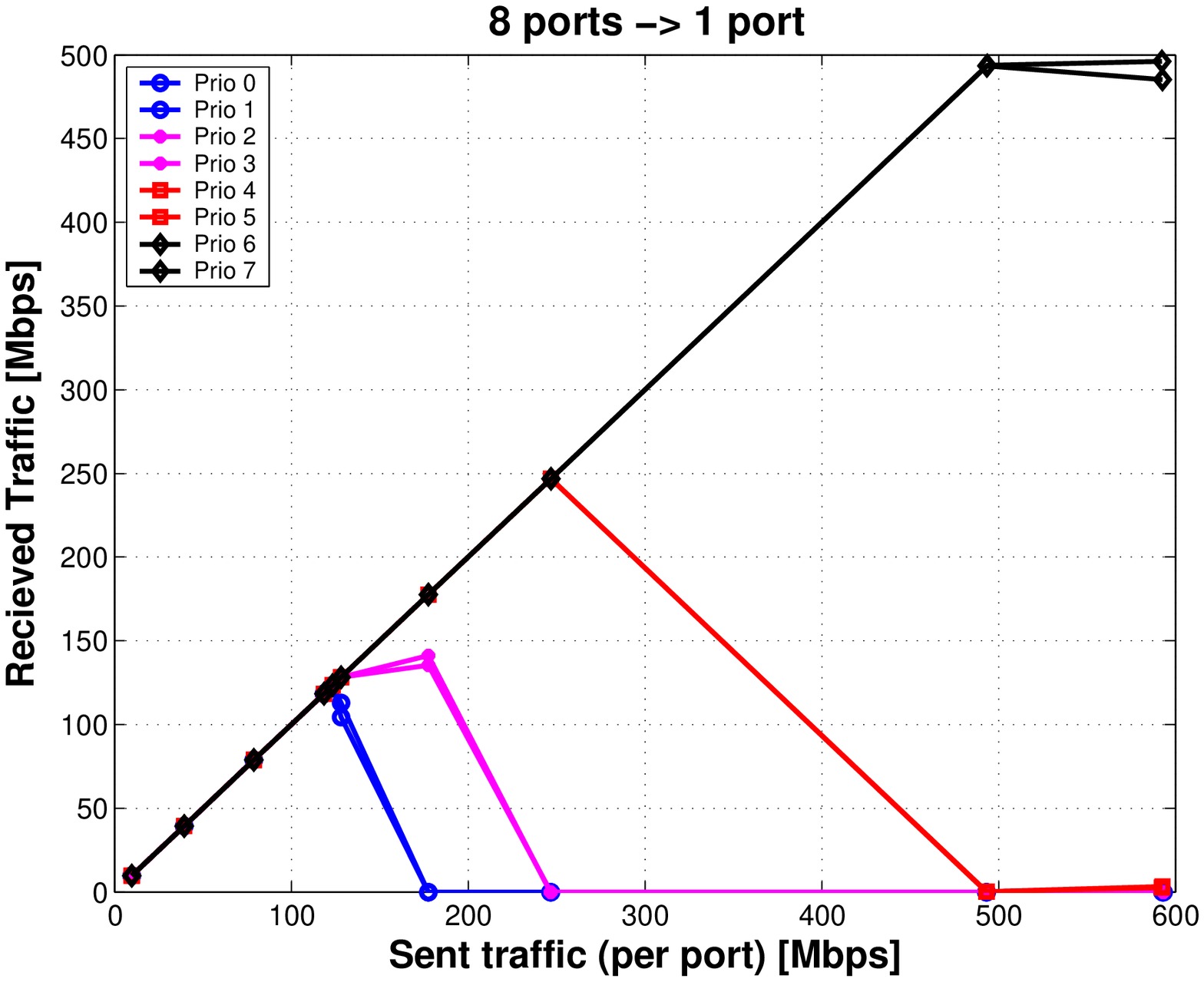}
    (a)
    \includegraphics[width=\columnwidth]{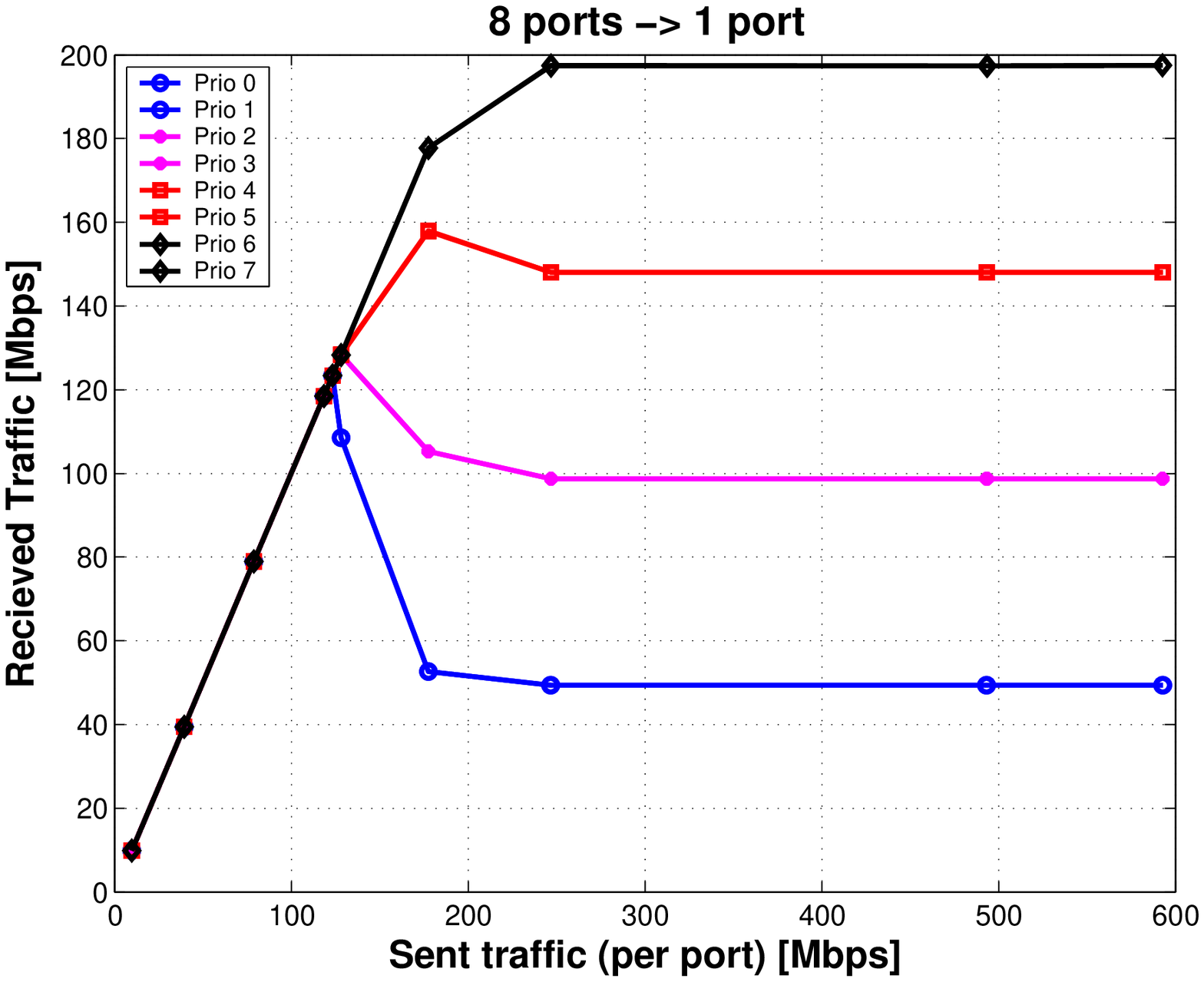}
    (b)
    \caption{QoS measurements:(a) strict priority algorithm
    (\textit{best-effort, normal, high, premium})(b) weighted
    round robin algorithm (weights: \textit{10, 20, 30, 40})}
    \label{fig:QoS}
\end{figure}

Figure~\ref{fig:QoS} presents the results of a QoS test. Eight GE ports send
constant bit rate (CBR) traffic (1518 bytes frames) to the same GE receiving
port. Each of the senders sets a different priority in the VLAN tag. Once the
intended throughput exceeds the line capacity the strict priority algorithm
gradually starves the lower priorities in favor of the higher ones. In the case
of the WRR algorithm the bandwidth is allocated to each priority proportionally
to its associated weight, once the line is saturated.

A possible use of QoS in the DataFlow is to assign a higher priority to the
control messages (like the messages from the L2SV to the DFM) with respect to
the main data flow.


\subsection{Ethernet Flow Control \label{sec:MeasFC}}
\begin{figure}
    \centering
    \includegraphics[width=\columnwidth]{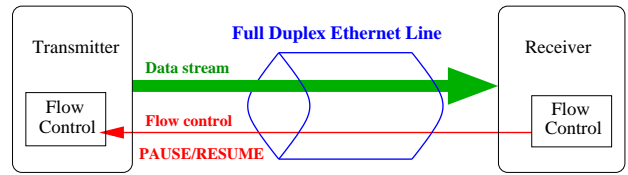}
    \caption{Ethernet Flow Control mechanism.}
    \label{fig:FCmechanism}
\end{figure}

On a full duplex Ethernet line, a slow receiver can limit the sending rate of a
fast transmitter, using Ethernet Flow Control Frames. The FC (Flow Control)
mechanism is presented in Figure~\ref{fig:FCmechanism}. When the receiver
becomes low in resources it sends a PAUSE frame, which will block the
transmitter. Once enough resources are available on the receive side a RESUME
(transmission) frame tells the transmitter to restart sending the data
stream.

A proper Ethernet Flow Control implementation guarantees no overflow occurs at
the level of Ethernet MAC buffers. Yet we must investigate Flow Control
propagation through the network switches, as well as inside the PC's (NIC,
Operating System (OS) and user-level application), in order to efficiently use
this feature in the ATLAS TDAQ system.

\subsubsection{Propagation through the switches}
\begin{figure}
    \centering
    \includegraphics[width=\columnwidth]{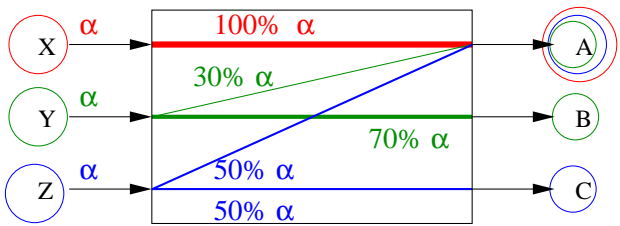}
    \caption{Flow control propagation testing setup.}
    \label{fig:FCThroughSwitch}
\end{figure}

In the setup illustrated in Figure~\ref{fig:FCThroughSwitch} all the
transmitters send CBR traffic at the same rate, denoted as $\alpha$.  X sends
100\% of $\alpha$ entirely to A, Y splits its traffic 30\% to A and 70\% to B,
while Z transmits 50\% $\alpha$ to A and 50\% to C. We gradually increase
$\alpha$ up to the line speed. When $\alpha$ becomes larger than 56\% of the
line speed the switch port from node A is oversubscribed. 

\begin{figure}
    \centering
    \includegraphics[width=\columnwidth]{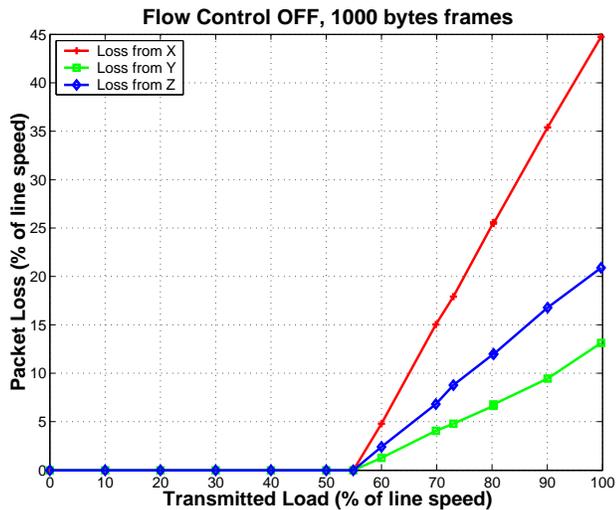}
    (a) 
    \includegraphics[width=\columnwidth]{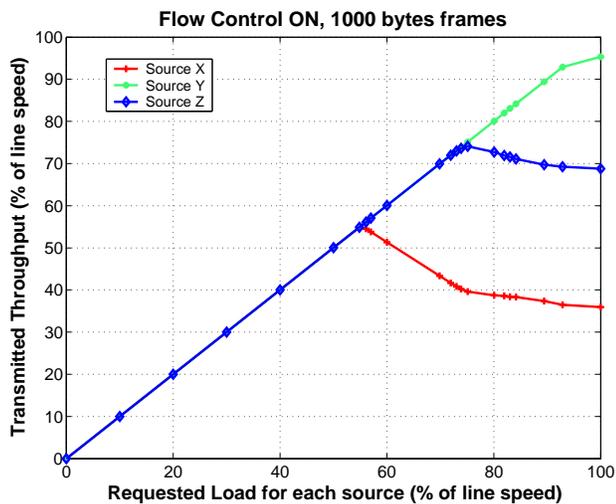}
    (b)
    \caption{Congestion analysis: (a) FC OFF, (b) FC ON}
    \label{fig:FCplot}
\end{figure}

When Flow Control is disabled on all the ports, packet loss occurs 
at the congested port (Figure~\ref{fig:FCplot} (a)). It is important to see if
the congested port affects the other traffic paths. Although some of the frames
directed to A are lost, B receives all the traffic from Y without any 
loss. The same observation is true for transmitter Z. This proves 
the input buffers have separate queues for each outgoing port,
therefore the switch is not susceptible to HOL (Head Of Line) blocking.

Once we enable Flow Control (both on the nodes and the switch's ports) we no
longer observe packet loss at port A. This proves the switch is propagating
Flow Control between its ports ensuring a lossless frame transfer between the
input and output switch buffers. The price paid for no packet loss is the
congestion spreading effect.  The congestion from port A causes a slowdown in
the transmission rate of all the ports that send frames to this destination
(X, Y and Z). Thus the sending rate from Y to B, as well as from Z to C will
be affected.  Figure~\ref{fig:FCplot} (b) demonstrates that the congestion
effect from port A spreads towards all the ports contributing to it
proportionally with the amount of traffic sent to the oversubscribed port.

The congestion spread caused by the choice of propagating Flow Control through
the switch has undesired effects in case of a node failure. We have modified
the A receiver to emulate a ``dead'' node with active FC\footnote{This may
happen if the Linux kernel crashes but the PC's NIC remains active, or if the
firmware crashes on a hardware device like the ROBs}. As the ``dead'' node
cannot empty the received frames from the low level MAC receive queue, it will keep
sending FC to the switch, as long as the latter tries to deliver packets to
node A. The congestion spreads, and all the transmitters are completely
blocked. No traffic reaches port B or C, as long as Y and Z keep sending
towards A. Therefore,  it will be important to detect and deal with
malfunctioning network nodes.

The choice of propagating Flow Control through the switches is manufacturer
depended. Some switches do not propagate FC. Packet loss occurs if the
congestion cannot be absorbed by the internal switch buffers. However, in this
case there is no congestion spreading effect.

\subsubsection{User level application -- traffic shaping} 
\begin{figure}
    \centering
    \includegraphics[width=50mm]{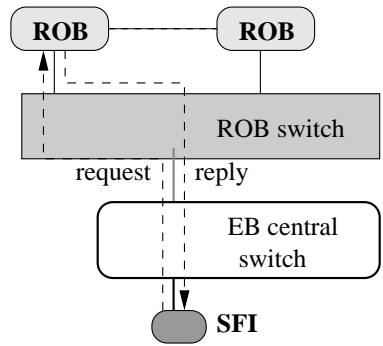}
    \caption{Request-response traffic Flow Control analysis.}
    \label{fig:FCTrafShape}
\end{figure}

One may think that
building a network with switches that propagate FC (and enabling it everywhere) guarantees
lossless communication between all the applications. This is not true if the
applications run on Linux OS PCs. 
We use one PC with a request-response program emulating the SFI functionality,
while the ROBs are emulated by modified traffic generators (see
Figure\ref{fig:FCTrafShape}).

In order to see if the PC sending rate can be reduced by Flow Control,
the SFI interrogates only one ROB, which is artificially
slowed down using a busy loop. The ROB cannot cope with the request rate,
asserts FC which propagates through the network, and slows down the PC's NIC.
The kernel sending queues will fill up and a call to \texttt{send} will fail.
If the \texttt{errno} (a global variable indicating the error
encountered while executing the \texttt{send} system call) value indicates no
buffer space was available for sending, the sending thread should sleep for
a while, then try to send again. Therefore the PC sending rate can be reduced
using Flow Control.

On the receive part the things are not that simple. If the PC's NIC or kernel
cannot cope with the received traffic rate, FC will be asserted by the NIC.
The problem appears when the user-level application cannot empty its
communication socket receive buffer. If the kernel cannot push frames to the
application's socket receive buffer, it will simply drop them without
any further notice. In other words, a lazy user level application is
not allowed to slow down (or even block) other peer processes which use
different communication sockets. On the other hand this kernel behaviour
limits the user-level application's ability to assert Flow Control when it
cannot cope with the incoming message rate.

Thus it is important that the DataFlow applications use a request-response
message flow scenario. Each node can take care not to request more than
it is prepared to receive: \textit{traffic shaping}.  The sending part of an application
is blocked if the number of outstanding request exceeds a certain threshold,
and becomes active once responses arrive back.

\subsection{Trunking (LAG - Link Aggregation Group) -- IEEE 802.3ad\label{sec:trunking}}

Trunking  allows grouping more physical links in order to form one
logical link with a higher bandwidth. The IEEE standard specifies that the order
of Ethernet frames shall aways be preserved by the trunk. Though, there's no
restriction for the allocation of the frames to the physical lines
within the trunk.

\begin{figure}
    \centering
    \includegraphics[width=70mm]{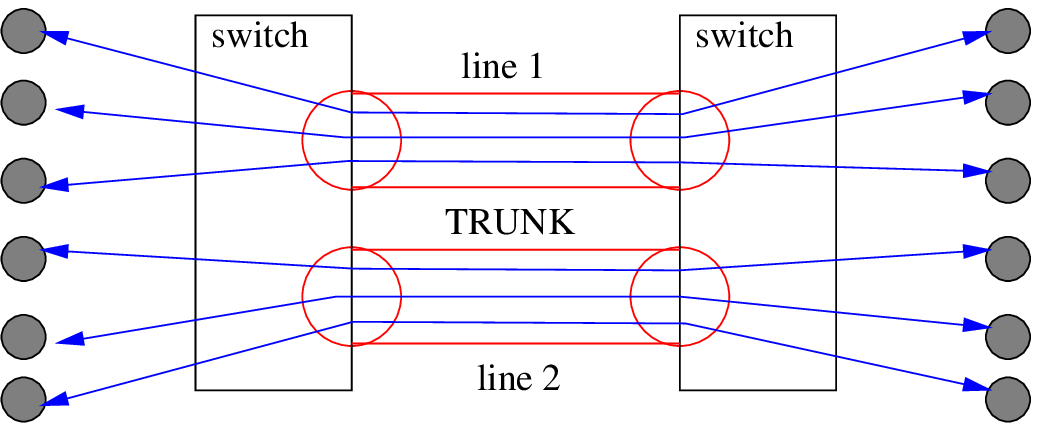} \\
    (a) \\
    \includegraphics[width=70mm]{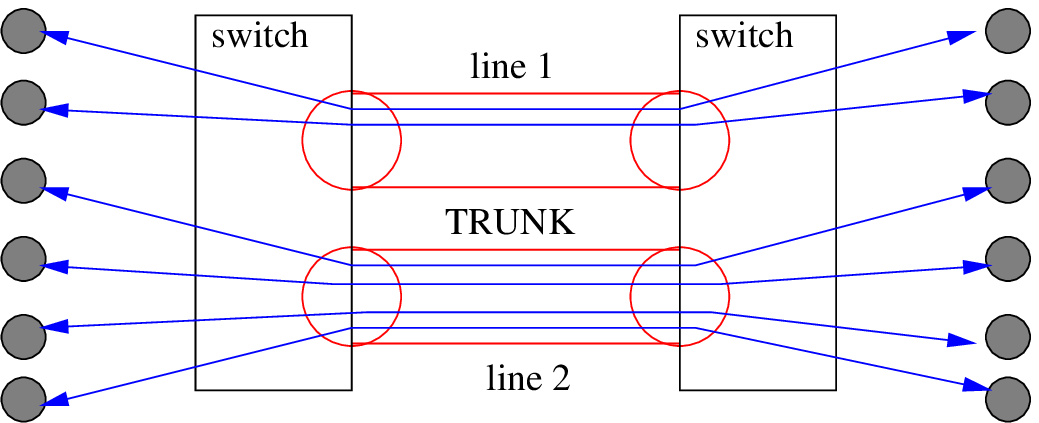} \\
    (b) \\
    \caption{Trunking: (a) even load balancing, (b) uneven load balancing}
    \label{fig:trunk}
\end{figure}

Once a physical link is associated to a pair of MAC addresses, all the frames
addressed to any of them will go on that link until the switch forgets about
one of the two MAC addresses (due to aging). This leads to a  potential load
balancing problem. Figure \ref{fig:trunk} (a) shows an even distribution of
the traffic, while Figure \ref{fig:trunk} (b) reveals an uneven distribution of
the traffic on the lines withing the trunk. 

We will further use the term ``connection'' to denote a pair of MAC addresses,
each one located on a different side of the trunk. The particular switches we
have tested randomly allocate the physical links (within the trunk) to a
connection, regardless of the traffic amount. In Figure \ref{fig:trunk} (b) it
may happen that the upper link of the trunk is not utilized, while the lower
one is oversubscribed.

From the ATLAS point of view, all the nodes from one side of the trunk will
require a similar bandwidth. Considering there will be a large number of
\textit{connection} pairs (there are approximately 1600 ROBs) across the
trunk, a random allocation algorithm will be effective, if we decide to use
trunks in the DataFlow network.


\section{Conclusions\label{Conclusions}}
Ethernet is considered the most suitable technology for the DataFlow network
because:
\begin{itemize}
    \item It satisfies the bandwidth requirements for the ATLAS DataFlow. 
    \item Segments with different speeds can be transparently
    interconnected via switches: 100 Mbps, 1 Gbps  and 10 Gbps.
    \item It is multi-vendor technology with long term support. 
    \item Ethernet is a commodity:
    \begin{itemize}
	\item The price of the GE UTP NIC is approximately 80 USD. However,
	most high-end PCs are equipped with an on-board GE NIC.
	\item The GE switch port cost is constantly dropping with time.
    \end{itemize}
    \item PCs have become fast enough to cope with the Gigabit Ethernet line
    speed.  A dual Pentium 4, with a CPU frequency of 2.4 GHz, can receive
    approximately 70 MBytes/second, in a request--reply traffic scenario for
    Event Building.
    \item Ethernet has an evolutionary upgrade path to high speed.
\end{itemize}

The DataFlow network must be a high performance network, as its nodes run
real-time applications. That's why the Ethernet features presented in this
article must be verified on every switch before its integration to the network. 


\begin{acknowledgments}
The authors wish to express their gratitude to:
\begin{itemize}
    \item Razvan Beuran and Mihail Ivanovici for the QoS measurements;
    \item Catalin Meirosu and Jamie Lokier for developing the GE network tester;
    \item Micheal LeVine, Jamie Lokier and Razvan Beuran for developing the FE
    network tester.
\end{itemize}
\end{acknowledgments}


\end{document}